\documentstyle[preprint,revtex]{aps}
\def\lsim{\kern .3em\raise1pt\hbox{$\scriptstyle\twiddle$}\kern
-.10em\raise2pt\hbox{$\scriptstyle<$}\kern .3em}

\def\ssc{\journal Solid State Communications, }

\def\lasr{$\rm La_{1.85}Sr_{.15}CuO_4$}

\def\lasrs{$\rm La_{1.85}Sr_{.15}CuO_4$\ }

\def\bisco{$\rm Bi_2Sr_2CaCu_2O_8$}
\def\biscos{$\rm Bi_2Sr_2CaCu_2O_8$\ }

\def\pr{\journal Phys. Rev., }

\def\prb{\journal Phys. Rev. B, }

\def\prl{\journal Phys. Rev. Lett., }

\def\rmp{\journal Rev. Mod. Phys., }

\gdef\journal#1, #2, #3, 1#4#5#6{               
    {\sl #1~}{\bf #2}, #3 (1#4#5#6)}            

\def\etal{{\it et al.}}

\def\pmb#1{\setbox0=\hbox{#1}%
  \kern-.025em\copy0\kern-\wd0
  \kern.05em\copy0\kern-\wd0
  \kern-.025em\raise.0433em\box0 }

\def\vep{\varepsilon}
\begin{document}
\draft
\rightline{cond-mat/9303027 \hfill NSF-ITP-93-25}
\begin{title}
Method for Measuring the Momentum-Dependent Relative Phase

of the Superconducting Gap of High-Temperature Superconductors
\end{title}
\author{M.E. Flatt\'e}
\begin{instit}
Institute for Theoretical Physics

University of California,
Santa Barbara, CA 93106-4030
\end{instit}
\moreauthors{S. Quinlan and D.J. Scalapino}
\begin{instit}
Physics Department

University of California,
Santa Barbara, CA 93106-9530
\end{instit}
\begin{abstract}
The phase variation of the superconducting gap over the (normal) Fermi
surface of the high-temperature superconductors remains a significant
unresolved question. Is the phase of the gap constant, does it change
sign, or is it perhaps complex?
A detailed answer to this question would provide important constraints
on various pairing mechanisms.
Here we propose a new method for
measuring the relative gap {\it phase\/} on the Fermi
surface which is direct, is angle-resolved, and probes the
bulk.  The required experiments involve measuring phonon
linewidths in the normal and superconducting state, with
resolution available in current facilities.  We primarily address the
\lasrs material, but also propose a more detailed study of a specific
phonon in \bisco.
\end{abstract}
\pacs{74.25.Kc, 74.72.Dn, 74.72.Hs}
\narrowtext

The momentum and frequency
dependence of the superconducting gap reflects the underlying
structure of the pairing interaction.
In traditional low-temperature superconductors the gap depends weakly
on momentum
with negligible anisotropy on the (normal) Fermi surface, and
exhibits a clear frequency dependence near important phonon energies.
This implies that the effective pairing interaction is short range in
space, retarded in time, and relatively isotropic.
Detailed analysis of the frequency dependence of the gap determined from
$I(V)$ measurements provides the most direct evidence that the
electron-phonon interaction\cite{BCS} is responsible for superconductivity in
these materials.
The {\it p-wave} gap
in the $^3$He superfluids\cite{Leggett}
and the evidence of nodes\cite{Broholm} consistent with
{\it d-wave} pairing
in some of the heavy fermion superconductors suggest
that the basic interactions in these systems repel at momentum
transfers comparable to the Fermi momentum.
The exchanges of paramagnetic\cite{Anderson}
or antiparamagnetic\cite{ScalapinoHirsch}
spin-fluctuations have been
proposed as models for such interactions.

A variety of gap symmetries have been suggested for the cuprate-oxide
superconductors based on particular models of the pairing mechanism.
Phonon and charge fluctuation mediated mechanisms typically lead to gaps
which may vary in magnitude but remain constant in phase over the
Fermi surface.
Here we will call such gaps {\it anisotropic s-wave} gaps.
Pairing interactions associated with the exchange of an
antiferromagnetic spin-fluctuation in the Hubbard model
produce a $d_{x^2-y^2}$ gap which
changes sign on the Fermi surface\cite{Bickers,Monthoux}.
Slave boson approximations and variational calculations for the {\it
t-J} model have led to $d_{x^2-y^2}$ and {\it complex d-wave}
gaps\cite{complexd}.
One example of a
{\it complex d-wave} gap has a
constant magnitude, but a phase which varies as
$\exp(2i\phi)$, where $\phi$ is the angular location of a point on the
Fermi surface.
Clearly, a method for determining the relative phase of the gap between
different points on the Fermi surface would substantially
constrain any theory of high-temperature superconductivity.
Here we show how neutron scattering measurements of phonon linewidths
can provide such a probe.

This method involves
locating and measuring
anomalies (discontinuities) in phonon linewidths which occur
as a function
of temperature or frequency at the quasiparticle-pair-production threshold.
Previously a method was
proposed\cite{Flatte} for measuring the angle-resolved gap {\it magnitude}
using only the threshold {\it frequencies}. That method has been
applied\cite{Flattebisco}
to recent data\cite{Mook} on
a phonon in \bisco.  Now we demonstrate that measuring the threshold
discontinuity
size (in addition to its onset frequency) extends that method to
produce angle-resolved measurements of the relative gap {\it phase}
between two points on the Fermi surface.

Other methods used to detect the existence of {\it d-wave} gaps search
for pronounced gap magnitude anisotropy.  The temperature-dependence of
various thermodynamic quantities may contain anomalous (power-law)
contributions from gap nodes. A direct probe of gap magnitude, angle-resolved
photoemission, measures the gap magnitude around the Fermi surface.
Rather than relying on gap magnitude nodes to indicate changes in the gap
phase, our method directly
determines the relative phase of the gap, $\Delta\phi$, between
different points on the Fermi surface. This type of measurement is
crucial for theory because, even
if the gap magnitude is isotropic, the phase could make it {\it d-wave}, such
as for the gap function $\exp(2i\phi)$.
Conversely, gap nodes do not guarantee gap sign changes.
The method we propose has adequate Fermi surface angular resolution
and, most importantly,
probes the {\it bulk} properties of the superconductor.

The method for measuring the energy-gap magnitude in Ref. \cite{Flatte}
took advantage of the following geometrical arguments. For phonon
momenta larger than an inverse coherence length, the
two quasiparticles produced from a phonon's decay
appear at only two places on
a quasi-two-dimensional
Fermi surface.
Fig.~\ref{FS} shows this geometry.  Both the head and the
tail of the phonon wavevector (two are shown, vectors $\bf q_1$ and
$\bf q_2$) must lie on the Fermi surface.  The quasiparticles are
created at the head of this vector, and at the Fermi surface point
opposite the tail (as shown in Fig.~\ref{FS} for $\bf q_2$, where the
two points are labeled $\bf k_o$ and ${\bf q_2} -{\bf  k_o}$).
Any other placement of the
phonon wavevector results in an energy cost from creating quasiparticles
off the Fermi surface which far exceeds the energy of a characteristic
phonon. The minimum energy for a phonon to decay into two quasiparticles
is thus a function of momentum.  This energy constraint defines a surface
in $({\bf q},\omega)$ space called the {\it threshold} surface.

A derivation similar to that in previous work on
three-dimensional superconductors with isotropic
gaps\cite{Bobetic} identifies a discontinuity in the
linewidth of a phonon when it crosses the threshold
surface. A crossing can occur either through variation of the phonon
frequency and momentum (moving along a dispersion curve) or altering the
threshold surface (changing the temperature).
In order to determine the gap magnitude, the size of the discontinuity
is unimportant, so long as it is observable.

We show in this Letter that the size of the discontinuity yields
a direct measurement of the phase of the gap. For a given phonon,
the ratio of the
discontinuity at low temperature to the normal-state linewidth is
\begin{equation}
{\Delta\gamma\over \gamma_N} =
{\pi\over 2}\cos^2\left({\Delta\phi\over 2}\right),\label{discon}
\end{equation}
where $\Delta\phi$ is the relative phase
between gaps at the locations where quasiparticles are created
(described above). To understand this, we note that the phonon linewidth
is proportional to the imaginary part of the
density-density response function\cite{MC}
\begin{eqnarray}
\hbox{Im}&&P({\bf q},\omega) = \label{bigpi}\\
&&{\pi\over N}
\sum_{\bf k} \Big\{ \left[1+ {\vep_{{\bf q}-{\bf k}}\vep_{\bf k} -
\hbox{Re}(\Delta_{\bf k}\Delta^*_{{\bf q}-{\bf k}})
\over
E_{{\bf q}-{\bf k}}E_{\bf k}}\right]
\, \left(f(E_{{\bf q}-{\bf k}})-f(E_{\bf k})\right)\delta(\omega
-E_{{\bf q}-{\bf k}}+E_{\bf k})\nonumber\\
&&\quad + {1\over2}
\left[1-{\vep_{{\bf q}-{\bf k}}\vep_{\bf k}-\hbox{Re}(\Delta_{\bf k}
\Delta^*_{{\bf q}-{\bf k}})
\over E_{{\bf q}-{\bf k}}E_{\bf k}}\right]
\, \left(f(E_{{\bf q}-{\bf k}})+f(E_{\bf k})-1\right)\delta(\omega-
E_{{\bf q}-{\bf k}}-E_{\bf k})\Big\}.\nonumber
\end{eqnarray}
Here $\vep_k$ is the normal state dispersion relation for electrons
(measured from the chemical potential), $f(\vep)$ is the Fermi function,
$\Delta_{\bf k}$ is the
(momentum-dependent) gap, and $E_{\bf k}=\sqrt{\vep^2_{\bf k} +
\Delta^2_{\bf k}}$.
The first term in Eq.~\ref{bigpi} arises from the scattering of thermal
quasi-particles, and the second term corresponds to quasiparticle
pair production.
The square brackets contain the usual BCS coherence factors.
When the phonon energy
exceeds the minimum energy $|\Delta_{\bf k_o}|+|\Delta_{{\bf q}-{\bf
k_o}}|$, pair production of quasiparticles occurs.
The discontinuity in the
linewidth arises due to the square-root divergence in the density of
states near each of the points $\bf k_o$ and ${\bf q}-{\bf k_o}$.
The coherence factor regulates the
magnitude of the discontinuity.
When both quasiparticles are created on the Fermi surface, the coherence
factor reduces to this simple form\cite{magnetic}:
\begin{equation}
{1\over 2}\left(1
+ {{\rm
Re}[\Delta_{\bf k_o}\Delta^{*}_{{\bf q}-{\bf k_o}}]\over
|\Delta_{\bf k_o}||\Delta_{{\bf q}-{\bf k_o}}| }
\right)= \cos^2\left({\Delta\phi\over
2}\right).\label{onFS}
\end{equation}

We now consider two different phonons in \lasr, shown
as $\bf q_1$ and $\bf
q_2$ on the Fermi surface\cite{Hybertson} (Fig.~\ref{FS}).
We will examine their behavior for
four different gap functions:

{\it i. isotropic.}
\begin{equation}
\Delta_{\bf k} = \Delta(T).
\end{equation}
We choose $T_c=40K$ and $\Delta(T)$ to follow the BCS temperature dependence
with
$2\Delta(0)=6k_BT_c$.
The minimum energy of decay for phonons in this
superconductor is not momentum dependent.

{\it ii. constant phase with nodes.}
\begin{equation}
\Delta_{\bf k} = {\Delta(T)\over 2}|\cos(k_xa) - \cos(k_ya)|\label{abdwave}
\end{equation}
has nodes on the Fermi surface
at the four locations indicated with black squares on
Fig.~\ref{FS}.  This is a limiting example of an {\it anisotropic
s-wave} gap.

{\it iii. $d_{x^2-y^2}$.}
\begin{equation}
\Delta_{\bf k} = {\Delta(T)\over 2}\left\{\cos(k_xa) - \cos(k_ya)\right\}
\label{dwave}
\end{equation}
has the same gap magnitude as the above {\it anisotropic s-wave} gap,
and thus is indistinguishable
from it in most probes. However, the relative phase
$\Delta\phi$ can take
two values: $0$ and $\pi$.  The former case
occurs for phonon $\bf q_1$, while the latter for $\bf
q_2$.

{\it iv. complex d-wave.}
\begin{equation}
\Delta_{\bf k} = \Delta(T) \exp({2i\phi})
\end{equation}
has the same gap magnitude as the isotropic case, and so is
indistinguishable from it for most probes. The relative phase
between two points, however, can take on a continuum of values
from $0$ to $\pi$.
We will focus on
the above four gaps in this Letter although the theory applies to all gap
functions.

The ratio of the linewidth
at a low temperature ($T_c/2$) to that in the normal state is shown for
all four gaps for
phonon
${\bf q_1}=(0.4\pi,0)$ in Fig.~\ref{phonon1} and ${\bf q_2}
= (\pi,0.5\pi)$ in Fig.~\ref{phonon2}.
This ratio depends on the imaginary part of the polarization (density-density)
response according to the following equation:
\begin{equation}
{\gamma({\bf q},\omega T_c/2) \over \gamma({\bf q},\omega,T_c)} =
{ {\rm Im}P({\bf q},\omega,T_c/2)\over
 {\rm Im}P({\bf q},\omega,T_c)}
.\label{linew}
\end{equation}
We find this ratio convenient because it does not depend on
the electron-phonon coupling.

There are two discontinuities in the linewidths for the {\it anisotropic
s-wave} and $d_{x^2-y^2}$ gaps
in Fig.~\ref{phonon1} because there are two possible ways to create
quasiparticles on the Fermi surface: the normal process, labeled by $\bf
q_1$ in Fig.~\ref{FS}, and an Umklapp process, the dotted vector on
Fig.~\ref{FS}.  Since the gap magnitudes
in these regions differ, the discontinuities occur
at different energies.  The relative phase for these two gap
functions is identical for both channels, so the linewidths are
indistinguishable on Fig.~\ref{phonon1}.

There is a single discontinuity for the
{\it isotropic}-gap case because both of these decay channels have the same
minimum energy, $2\Delta(T)$.
For the {\it complex d-wave} gap,
the gap magnitudes for these two decay channels
are the same, but the phase of the gap at $\bf k_o$ and $\bf q_1\rm -
\bf k_o$ differ. For the normal
case, the phase difference
$\Delta\phi = .35\pi$, while for the Umklapp case, $\Delta\phi = .22\pi$.
These relative phases yield
discontinuities of $73\%$ and $88\%$ respectively
of the {\it isotropic}-gap result.
Since they occur at the same energy, the observed discontinuity has
an intermediate size of $\sim 80\%$, as
found in Fig.~\ref{phonon1}.

For the momentum transfer $\bf q_2$, shown in
Fig.~\ref{phonon2} there is only one decay channel, so there is a
single discontinuity for all four gap functions. For the constant phase
gaps, {\it (i)} and {\it (ii)}, this discontinuity has its maximum size. For
the $d_{x^2-y^2}$ gap, the relative phase between $\bf k_o$ and $\bf
q_2$ is $\pi$, so there is no step in the linewidth. There is merely a
discontinuity in the slope of the linewidth.  For the {\it complex
d-wave} gap, the relative phase is quite close to $\pi$, and the size of
the discontinuity reaches merely $2\%$ of maximum.
It is clear in Figs.~\ref{phonon1} and \ref{phonon2}
that not only are superconductors with different gap magnitudes
distinguishable from each other ({\it isotropic} from {\it anisotropic
s-wave}),
but  those which only differ by their gap phase as well ($d_{x^2-y^2}$
from {\it anisotropic s-wave} and {\it complex d-wave} from {\it isotropic}).

Figs.~\ref{temp} and~\ref{temp2} shows the
temperature-dependence of phonons with
momenta $\bf q_1$ and $\bf q_2$, respectively,
and energy $\omega=\Delta(0)$ in the four superconductors.
Although the energies of phonons with momentum $\bf q_2$
have not been
measured, a candidate phonon should exist\cite{phonons}
for a large range of values of $\Delta(0)$. There are three known candidates
for $\bf q_1$.

In Fig.~\ref{temp}
all of the superconductors have an enhanced linewidth below the
critical temperature ($T_c$),
and the discontinuities occur at a temperature resolvable from
$T_c$.
For the gap functions with anisotropic gap magnitudes, the two
decay channels are resolvable as
{\it separate} discontinuities.
A linewidth  enhancement at temperatures just above the discontinuity,
similar to those shown,
was observed in several phonons in a classic
niobium experiment\cite{Shirane}.
The enhancement
for the {\it complex d-wave} gap is lower than the {\it isotropic}
one
because the relative phase between $\bf k_o$ and ${\bf q_1}-{\bf k_o}$
differs from $0$.  From Eq.~\ref{discon}, there will be an
enhancement before the discontinuity
unless the relative phase exceeds $.41\pi$.

A relative phase of $\pi$ between
$\bf k_o$ and ${\bf q_2}-{\bf k_o}$ in Fig.~\ref{temp2} for the
$d_{x^2-y^2}$ case
causes the linewidth to
monotonically
decrease from $T_c$ without a discontinuity.
For the {\it complex d-wave}, the
relative phase is close to $\pi$, causing the
linewidth at $T_c$ to drop, and the discontinuity
at low temperature to be extremely small.
Thus the discontinuity size,
and the slope of the linewidth
at $T_c$ probe the relative phase of the gap.

Recently Mook \etal\cite{Mook}
reported the narrowing of a
low-energy phonon in \biscos upon cooling through $T_c$.
The resolution of that measurement
was inadequate to determine the structure around the discontinuity. However,
performing the measurement at a slightly higher resolution
should determine the nature of that structure.
The enhancement prior to the discontinuity
expected for {\it s-wave} gaps would not be
expected for this phonon if the superconductor had a
$d_{x^2-y^2}$ gap.

Other electronic contributions to a phonon's decay rate
besides the pair processes could complicate the interpretation of results.
We have described how, if the relative phase is small, for temperatures
slightly above the discontinuity the linewidth
will increase as the temperature drops. If,
however, the relative phase is close to $\pi$, the linewidth
decreases just before the discontinuity. Since other contributions to
the linewidth are expected to be smooth near the pair-production
threshold, even if those other contributions are large,
the temperature dependence of the linewidth
just above the discontinuity will retain this strong dependence on the
relative gap phase.

The probe described in this Letter provides a direct,  angle-resolved
measurement of the relative gap phase in high-temperature superconductors'
bulk.
Such measurements can constrain the range of acceptable theories
of high-temperature superconductors.

We thank B.-H. Sch\"uttler for suggesting an effective way of
numerically calculating the polarizability. S.Q. and D.J.S. acknowledge
the support of the National Science Foundation grant DMR90-02492.
M.E.F. acknowledges the support of this work by the
National Science Foundation under Grant No. PHY89-04035. The numerical
calculations reported in this paper were performed at the San Diego
Supercomputer Center.

\figure{Fermi surface of \lasr, taken from the parametrization of
Ref.~\cite{Hybertson}. Nodes in the
$d_{x^2-y^2}$ gap function
occur on the Fermi surface at the black squares. The sign of
the $d_{x^2-y^2}$ gap is indicated on the Fermi
surface.
This Letter examines in detail
phonons with momenta ${\bf q_1}=(0.4\pi,0)$ and ${\bf q_2}=(\pi,0.5\pi)$.
The two quasiparticles
created by the decay of phonon $\bf q_2$ form at locations
$\bf k_o$ and ${\bf
q_2}-{\bf k_o}$ on the Fermi surface. There are two decay channels for
phonon $\bf q_1$, the normal, labeled $\bf q_1$, and Umklapp (dotted
vector).\label{FS}}
\figure{The linewidth of phonons with momentum $\bf q_1$ as a function
of frequency for four different gaps. In the remaining figures, results
for these four cases will be plotted as follows: {\it isotropic}
(solid), {\it anisotropic s-wave} (dashed), $d_{x^2-y^2}$ (dotted), and
{\it complex d-wave} (dot-dash). In this figure, the curve for the
$d_{x^2-y^2}$ gap lies on top of the {\it anisotropic s-wave} curve.
The abscissa is normalized to the normal linewidth at $T_c$.
\label{phonon1}}
\figure{As in Fig.~\ref{phonon1}, but for momentum $\bf q_2$. In
contrast to Fig.~2, here the $d_{x^2-y^2}$ curve is distinguishable
from the {\it anisotropic s-wave}.\label{phonon2}}
\figure{The linewidth of a phonon with momentum $\bf q_1$ and energy
$\Delta(0)$ as a function of
temperature for the four different gaps. As in Fig.~2, the dotted line
($d_{x^2-y^2}$ gap)
is not visible, since it lies on top of the dashed line ({\it
anisotropic s-wave}).\label{temp}}
\figure{As in Fig.~\ref{temp}, but for momentum $\bf
q_2$. Here the $d_{x^2-y^2}$ curve is substantially different from the
{\it anisotropic s-wave} curve.\label{temp2}}

\begin{references}
\bibitem{MC} Numerical calculations of $P({\bf q},\omega)$ for the
Hubbard model with $U/t=4$ find that ${\rm Im} P({\bf q},\omega)$ is
proportional to the $U=0$ result when $\omega/t$ is small (N. Bulut,
private communication).
\end{references}
\end{document}